\documentclass{article}

\usepackage{arxiv}

\usepackage[utf8]{inputenc} 
\usepackage[T1]{fontenc}    
\usepackage{hyperref}       
\usepackage{url}            
\usepackage{booktabs}       
\usepackage{amsfonts}       
\usepackage{nicefrac}       
\usepackage{microtype}      
\usepackage{lipsum}		
\usepackage{graphicx}
\usepackage[square,sort,comma,numbers]{natbib} 
\usepackage{doi}

\usepackage{amssymb}
\usepackage{hyperref}
\usepackage{tikz}
\usepackage{amsmath}

\usepackage{url}
\usepackage{float}

\usepackage{tabularx}
\usepackage{rotating}
\usepackage{booktabs, multirow}
\usepackage{soul}
\usepackage[normalem]{ulem}
\usepackage{xspace}
\usepackage{amsfonts}
\usepackage{cleveref}
\usepackage{enumitem}
\usepackage[switch,pagewise]{lineno} 

\newcommand{\GView}{GView\xspace}
\newcommand{\gvcomp}[1]{\textsf{#1}}  

\title{GView: A Versatile Assistant for Security Researchers}

\date{April 12, 2024}	

\author{ 
    {\hspace{1mm}Raul Zaharia} \\
	Al. I. Cuza University \& Bitdefender \\
	Iași, Romania \\
	\texttt{rzaharia@bitdefender.com} \\
	\And
	{\hspace{1mm}Dragoș Gavriluț} \\
	Al. I. Cuza University \& Bitdefender \\
	Iași, Romania \\
	\texttt{dgavrilut@bitdefender.com} \\
    \And
	{\hspace{1mm}Gheorghiță Mutu} \\
	Al. I. Cuza University \& Bitdefender \\
	Iași, Romania \\
	\texttt{gheorghitamutu@bitdefender.com} \\
    \And
	{\hspace{1mm}Dorel Lucanu} \\
	Al. I. Cuza University \\
	Iași, Romania \\
	\texttt{dorel.lucanu@gmail.com} \\
}

\renewcommand{\headeright}{ }
\renewcommand{\undertitle}{ }

\hypersetup{
pdftitle={GView: A Versatile Assistant for Security Researchers},
pdfsubject={cs.SE},
pdfauthor={Raul Zaharia, Dragoș Gavriluț, Gheorghiță Mutu, Dorel Lucanu},
pdfkeywords={Cybersecurity \and Automatic artifact identification \and Intuitive views \and Coherent data correlation \and Malware analysis},
}

\begin{document}
\maketitle

\begin{abstract}
Cyber security attacks have become increasingly complex over time, with various phases of their kill chain, involving binaries, scripts, documents, executed commands, vulnerabilities, or network traffic. 

We propose a tool, \GView, that is designed to investigate possible attacks by providing guided analysis for various file types using \emph{automatic artifact identification, extraction, coherent correlation\,{\rm\sl\&},inference, and meaningful\,
{\rm\sl\&}\,intuitive views at different levels of granularity} w.r.t. revealed information.
The concept behind \GView simplifies navigation through all payloads in a complex attack, streamlining the process for security researchers, and increasing the quality of analysis. \GView is generic in the sense it supports a variety of file types and has multiple visualization modes that can be automatically adjusted for each file type alone. Our evaluation shows that \GView significantly improves the analysis time of an attack compared to conventional tools used in forensics.
\end{abstract}

\keywords{Cybersecurity \and Automatic artifact identification \and Intuitive views \and Coherent data correlation \and Malware analysis}

\section{Motivation and significance}
\label{sec:motivation}

\subsection{Motivation}
With over 1.2 billion recorded malware files this year~\cite{av-test-malware-stats}, cyber security attacks have become increasingly complex. These attacks rely on multiple payloads across different stages of the kill chain. The kill chain, as described by MiTRE\cite{mitre-matrix}, outlines techniques mapped to specific steps (\textit{recognition}, \textit{initial access}, \textit{execution}, \textit{persistence}, \textit{privilege escalation}, etc.), accomplished through various payloads written in different programming languages. These concepts are important for forensics investigations, where security researchers face the complex task of analyzing breach evidence and understanding the diverse payloads used in the kill chain, often requiring knowledge of specific forensics tools. These tools include binary disassemblers for different architectures, script deobfuscators, content extractors, image visualizations, network traffic analyzers, and metadata extractors for various file formats (often useful to identify possible hints that might indicate the purpose of a specific payload) as shown in Table~\ref{tab:file_types}. When it comes to complex attacks, the number of tools used for analysis can be extensive and often involves interconnectivity between them (data extracted from one tool is used with another), which the security researcher must manually handle.

Furthermore, access to breach evidence might not always be possible in hard copy, where data needing analysis is copied and transmitted to the investigation team. When dealing with sensitive data, the forensics is often performed on-site, providing direct access to the affected asset. In such scenarios, a security researcher might be required to download the necessary tools. 
In this context, we have identified a couple of problems that a forensics engineer face when analyzing~a~breach:
\begin{enumerate}
    \item a large number of tools that need to be learned and used;
    \item correlation and the consistency of information supplied by different tools;
    \item a lack of guidance in goal-oriented security research;
    \item a set of concerns related to used tools: Do they work locally? Are they sending any kind of data to an external service? Can they work on different operating systems?
\end{enumerate}
A possible solution to these problems, as well as a way to ease the work of a security analyst, is to develop only one analysis-assistant tool that has to:
\begin{itemize}[itemsep=1pt,topsep=2pt,partopsep=1pt]
    \item be able to automatically extract relevant artifacts and display them in an intuitive way;
    \item soundly correlate the information supplied by different specialized components;
    \item infer new information based on what was revealed until that moment;
    \item have meaningful\,\&\,intuitive views at different levels of granularity w.r.t. revealed information;
    \item guide the analyst by supplying hints;
    \item perform the analysis locally, instead of sending the data to a cloud processor; this is, in particular, important when it comes to sensitive data;
    \item be available for most common operating systems (Windows, Linux, Mac);
    \item run through an SSH connection, especially for analyzing attacks involving IoT devices like the Mirai botnet\cite{mirai-usenix}, which require direct analysis on the device (e.g., routers).
\end{itemize}

Starting from these requirements, we developed \GView (\underline{\textsf{{G}}}eneric \underline{\textsf{{View}}})\footnote{\url{https://github.com/gdt050579/GView}}, a tool specifically designed to aid forensic engineers and security researchers. 

\GView is developed using AppCUI\footnote{\url{https://github.com/gdt050579/AppCUI}}, a TUI (Text/Terminal User Interface) open-source library that we have developed to be compatible with various operating systems and terminals, including the possibility of working through SSH connections. 

\GView is a complex tool and its design and development raised non-trivial software engineering challenges, some of which are described in Section~\ref{sec:arch}. \GView is available on GitHub\footnote{\url{https://github.com/gdt050579/GView}} and can be used under the MIT license.

\subsection{Use cases}
\label{sec:use-cases}
\GView was designed with 3 major use cases in mind:
\begin{enumerate}
    \item Static analysis for \textbf{forensics investigations}. Static analysis plays an important role in forensic investigations. \GView is designed to understand a vast majority of file types, thus limiting the need for additional tools, as it offers an all-in-one package. \GView also operates locally, ensuring data processing without relying on external services in any cloud processing. Additionally, it seamlessly runs on multiple operating systems, making it highly valuable for on-site forensic tasks.

    \item Usage in \textbf{Security Operations Center} (SOCs). A key attribute of a SOC team is that it has to respond rapidly during an ongoing attack. When presented with a collection of artifacts (files, event logs, etc.), it must promptly decide if  an attack is in progress. \GView offers various automated analysis features, including precise string identification (such as paths, IPs, and registry keys), as well as guided functionalities like script deobfuscation and disassembly. These capabilities assist in the decision-making process, enabling SOC teams to act promptly.
    
    \item \textbf{In-lab usage}. \GView incorporates a robust plugin system that enables the addition of diverse data identification plugins and smart viewers. This feature allows security researchers to extend \GView with customized plugins that could assist in adding detection rules/signatures supplied by another security product. These plugins can access all the extracted data within \GView, including selections, buffers, text representations, and disassembly code.  It is worth noting that \GView operates under the permissive MIT license, which allows for unrestricted extensions and commercial utilization of the platform.
\end{enumerate}

The screencast~\cite{gview-screencast} and the scenario from Section~\ref{sec:example} are applicable for all of the above three use cases.

\subsection{Related Work}
\label{sec:related_work}
The Cyber Security landscape has prompted the development of numerous tools and technologies aimed at addressing the continuously evolving threats.
Malware analysis is a vital approach in this field that involves dissecting and studying malicious software to understand its behavior, capabilities, and potential impact.
In order to speed up the malware analysis process, an approach is to leverage existing tools, scripts, and libraries for common analysis tasks and automatize this process as much as possible.

\textbf{Network Forensics} tools such as Wireshark~\cite{wireshark,Wireshark2,Wireshark3} and NetworkMiner~\cite{NetworkMiner}
enable investigators to capture, analyze, and reconstruct network traffic to identify malicious activities, unauthorized access, and communication patterns between systems. \GView currently offers support for HTTP streams only (by {PCAP} plugin) and the plan is to extend it in the next future with DNS, ICMP, and several others. 
The information obtained from standalone network Forensics tools must be correlated with artifacts supplied by other tools. 
Using \GView, a security analyst can easily recompose the data into artifacts that she/he can easily analyze without leaving the session; then she/he can use additional plugins while keeping track of all hierarchical history of each artifact (e.g., packages that compose the stream, origin of the file, resources found in file) in an organized manner.

\textbf{Static Analysis} involves analyzing binary code, file structure, and embedded resources to identify suspicious patterns, signatures, and potential vulnerabilities. 
The state-of-the-art frameworks that are doing a good job here include IDA Pro \cite{wong2018mastering}, Ghidra~\cite{ghidra},
Binary Ninja~\cite{binary-ninja},
Hiew \cite{wong2018mastering}, or Radare2 \cite{wong2018mastering}.
\GView includes plugins that parse, extract indicators, analyze, and disassemble MZPE/ELF/Mach-O binaries in order to obtain features similar to those available in the above-mentioned frameworks. 

\textbf{Digital forensics} is essential in the investigation and evidence for incident response purposes. In this context, specialized tools play a crucial role in analyzing artifacts specific to the Windows operating system. For example, LECmd (LNK parser), as mentioned in \cite{PECmd_LECmd} and \cite{LECmd}, allows investigators to examine possible exploits or persistence mechanisms used by malicious programs by parsing LNK files. Similarly, PECmd (Prefetch parser), mentioned in \cite{PECmd2}, \cite{PECmd3}, and \cite{PECmd4}, provides the ability to analyze recently executed binaries by reading their dependencies and loaded modules from Prefetch files.
\GView, on the other hand, takes a comprehensive approach by encompassing the functionalities of these specialized tools. 

\textbf{Similar tools}. We identified forty-one tools that include similar functionalities with \GView and 
the conclusions of a comparison between these tools and \GView is given in \Cref{sec:impact}.

\section{Software description}
\label{sec:description}

\subsection{Software architecture}
\label{sec:arch}

At its core, \GView is built upon two major components: \textit{data identifiers},  \textit{smart viewers}, a \textit{coherent collaboration mechanism}, and \emph{guided security analysis}. 

A \textbf{data identifier} is a plugin that can analyze a buffer (either in its binary format or translated to a text representation) and automatically extract meaningful (specific information) from that buffer based on its type. Each such plugin has dedicated panels to display type-specific details. For example, the plugin for a Portable Executable\footnote{\url{https://learn.microsoft.com/en-us/windows/win32/debug/pe-format}} ({PE}) file uses these panels to present imports, exports, resources, version information, icons, sections, and other binary-specific details. Currently, \GView includes 20+ data identifiers for various types of \textit{text files}, \textit{network captures} ({PCAP}), various types of \textit{images} (e.g., {BMP}, {ICO}), various types of \textit{executable files} (e.g., {PE}, {ELF}, {MACHO}), and various types of \textit{archives} (e.g., {ZIP}, {ISO}); The current list is in Appendix \ref{appendix:gview_data_identifiers} and an up-to-date list can be found on \GView Github repository\footnote{\url{https://github.com/gdt050579/GView/tree/main/Types}}.

\textbf{Smart viewers} are components that represent the data in different ways. A {data identifier} plugin typically selects multiple smart viewers and designates one as the primary viewer. Users can easily switch between these viewers to select the visualization method that best suits their requirements. For example, a Windows ZIP self-extract binary is essentially a {PE} file, and as such, the selected {data identifier} plugin for this file would be {PE}. For this particular case, the {PE} plugin will choose three smart viewers: 
\begin{itemize} 
    \item a viewer that shows the binary content of the file;
    \item a viewer that supports navigating through the binary code disassembly;
    \item a viewer for accessing the ZIP content.
\end{itemize}
As the self-extract code holds no significant interest (it is the same for all Windows zip self-extract binaries), the last of the three viewers become the default primary viewer.

Each smart viewer provides a customizable interface for the data identifier plugin, enabling meaningful representation of the displayed information. This interface varies for each smart viewer and includes configuration options and callbacks. For example, the {buffer viewer} can be configured to highlight specific data offsets, while the {lexical viewer} can use callbacks (provided by the data identifier plugin) to aid the de-obfuscation process. Additionally, {smart viewers}  may infer some information by themselves (without the help of the data identifier plugin). For instance, the {buffer viewer} can identify string representations (e.g. Unicode) and numeric values (e.g., float/double in IEEE 754 format), while the lexical viewer can automatically hide comments in scripts/programs, which are commonly used to obfuscate code. These capabilities justify the classification of these viewers as being "smart".

\begin{figure}[t]
\vspace{-2ex}
\centerline{\includegraphics[scale=0.58,interpolate]{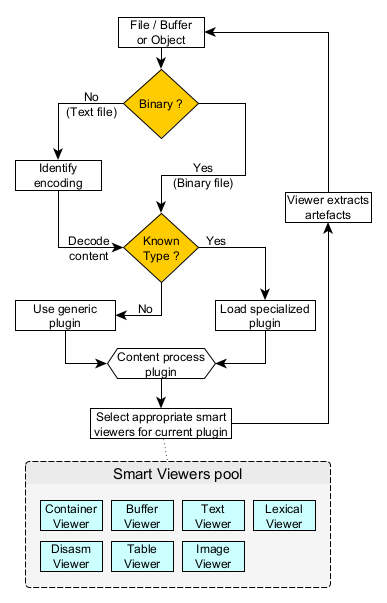}}
\vspace{-2.5ex}
\caption{\GView architecture overview}
\vspace{-3ex}
\label{fig:gview_flow}
\end{figure}

The iterative process is driven by a \textbf{coherent collaboration mechanism}, which correlates the artifacts extracted from previous steps. This means that actions like deobfuscating JavaScript code may result in additional snippets of code, which in turn requires further analysis and processing in order to understand their behavior.

The interconnection between \GView's {data identifiers} and {smart viewers} can be summarized by the following four steps, which are followed whenever a file/buffer or object is opened for analysis
(as showed by Figure \ref{fig:gview_flow}):
\begin{enumerate}
    \item Check if the content is binary or text. If the content is a text, try to understand the encoding and decode it to a common format (UTF-16).
    \item Search through all data identification plugins and find one that can interpret the data. If no such plugin is found, use a generic plugin that only takes into consideration if a file is binary or text.
    \item The data identification plugin selects one or multiple smart viewers that are appropriate for the current data. The user can switch between these plugins and can use them to extract artifacts or data from the existing content. 
    \item For the extracted components, steps 1 to 3 are repeated.
\end{enumerate}

Smart viewers and data identifiers often \emph{work together in order to achieve a high degree of collaborative, organizational, and structured investigation} on one or multiple samples at once, implementing communication between these two entities and a simple-to-use overview for navigation purposes.
These mechanisms can be organized into four kinds of bidirectional communications:
\begin{itemize}[itemsep=1pt,topsep=2pt,partopsep=1pt]
    \item A data identifier can communicate with multiple smart viewers showing the same data.

    \item For embedded artifacts to be extracted and analyzed, a collaboration between a data identifier with its current smart viewer is required. The result is an automatic artifact identification organized into an easy-to-use tree structure.

    \item Multiple smart viewers can communicate between them to exchange information about selections/positions and synchronize themselves.

    \item Smart viewers of the same type, but for different files, can communicate and synchronize between them to provide a way to find differences and similarities between multiple files.
\end{itemize}

\textbf{Guided security analysis} works as a complementary method with the collaboration mechanisms and it relies on information extracted with the following components: 

    Data identifiers, which focus solely on specific data format and structure, \emph{highlights type-specific information that might indicate a malicious intent}. 

    Plugins specific to data identifiers \emph{guide the analysis to a way of applying post-processing actions} on a given identified buffer. For example, on JavaScript data identifier the user can concatenate multiple strings that are being added, and propagate constant strings or reverse strings, all serving as deobfuscation mechanisms available to a researcher.

    Generic plugins, which are not bound to a specific data format, \emph{emphasize metadata, embedded artifacts, or possible hints from raw data} extracted by data identifiers or from data actively selected on a viewer by the analyst. The analyst \emph{can use them as clues} to what the suspicious sample does, what it seeks, how or to what it connects (or communicates), and under what circumstances.
    This is achieved by applying various algorithms (hashing, decompression) or post-processing actions (e.g., example attaching a class -- virtual coin wallet, registry, paths -- to certain types of strings, along with their prevalent usage).

\subsection{Software functionalities}
\label{sec:funct}

\GView is designed to smartly assist a security researcher with the following functionalities:
\begin{itemize}[itemsep=1pt,topsep=2pt,partopsep=1pt]
    \item automatically extracting relevant artifacts and intuitively displaying them;
    \item soundly correlating the information supplied by different specialized components;
    \item inferring new information based on what was revealed until that moment;
    \item supplying meaningful\,\&\,intuitive views at different levels of granularity w.r.t. revealed information;
    \item guiding the analyst by supplying hints;
    \item performing the analysis locally, instead of sending the data to a cloud processor; this is, in particular, important when it comes to sensitive data;
    \item to be available for most common operating systems (Windows, Linux, Mac);
    \item running through an SSH connection, especially for analyzing attacks involving IoT devices like the Mirai botnet\cite{mirai-usenix}, which require direct analysis on the device (e.g., routers).
\end{itemize}

\section{Illustrative examples}
\label{sec:example}

\label{sec:UseCase}
Here we discuss a relatively complex attack and see how \GView can help a forensics engineer. This scenario can also be viewed in the GView Screencast~\cite{gview-screencast}. Let us consider a hypothetical person (named Bob) who needs to install a Firefox browser into his server but ends up with ransomware. In terms of the attack, this will unfold in the following way:
\\
\textbf{Step 1.} Bob searches the internet for a Firefox browser download and chooses the wrong download link (not the official one from Mozila's website).
\\
\textbf{Step 2.} The fake website is based on a phishing attack (it mimics the looks of the original Mozilla website), and the download is not the actual Firefox installer.
\\
\textbf{Step 3.} Bob downloads a fake Firefox installer ("safe\_mozilla\_\-installer.exe") and runs it.
\\
\textbf{Step 4.} To avoid any potential detection from a security product, the downloaded file "safe\_mozilla\_installer.exe" is a ZIP self-extract password-protected archive. As such, upon execution will ask Bob for a password to continue the process.
\\
\textbf{Step 5.} The password was provided on the website but as a picture instead of the text (again, to avoid any automated detection system that might have extracted the password and used it to scan the content of the archive).
\\
\textbf{Step 6.} Bob types the password, and the content of the archive is extracted and executed. The content is actually ransomware that encrypts his critical documents from his storage.

With this in mind, let us see how \GView could help us in a forensics investigation. We assume that Bob's computer is behind a network appliance that records the entire traffic that goes in and out of Bob's endpoint. That recording, exported as a PCAP file, will be the starting point in this case.

When opening the PCAP file, \GView automatically identifies all connections made in and out of Bob's PC. However, since there are so many, and knowing that in the end ransomware was deployed, we can safely assume that a JavaScript code was responsible for this download. \GView has a useful mechanism where we can type a string (in the PCAP viewer, e.g., an extension) and it highlights all connections that contain that string. For our case, we just need to use ".js" string to identify a file called "analytics.js" that could be relevant (Figure~\ref{fig:Use-Case-Analytics,js}).

\begin{figure}[htbp]
   \centering
    \includegraphics[width=0.9\textwidth]{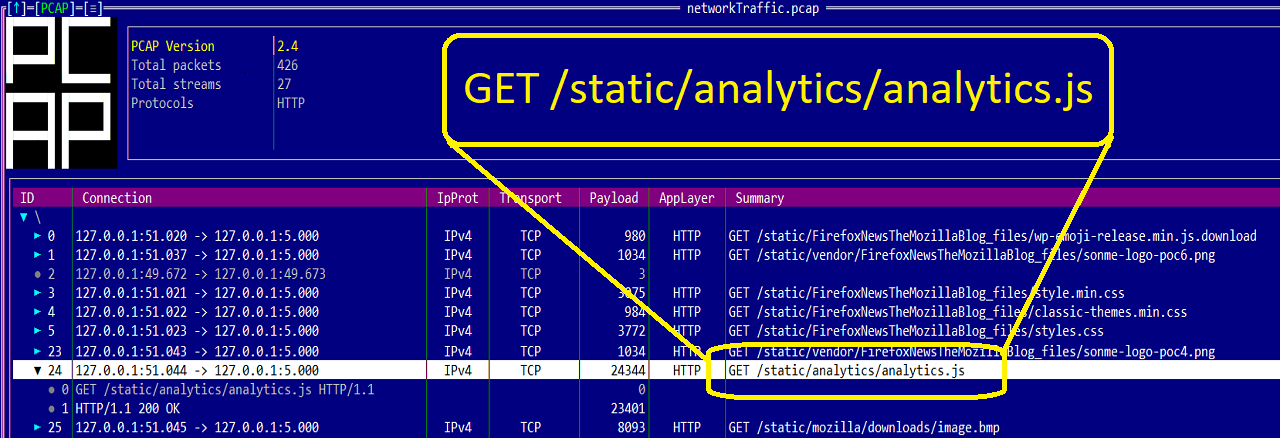}
    \vspace{-1ex}
    \caption{JavaScript components from the network traffic}
    \label{fig:Use-Case-Analytics,js}
    \vspace{-1ex}
\end{figure}

\begin{figure}[htbp]
   \centering 
    \includegraphics[width=0.9\textwidth]{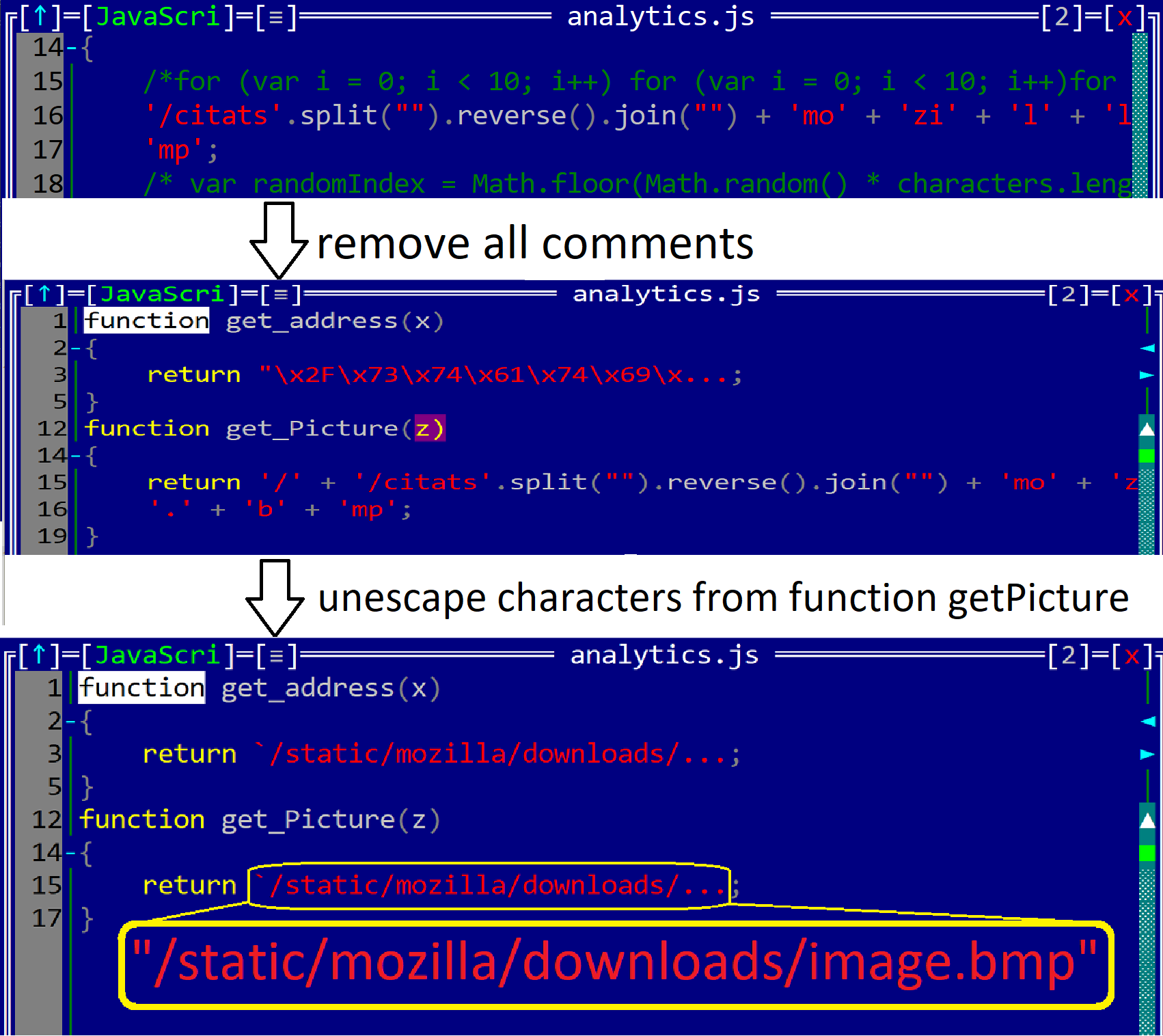}
    \vspace{-1ex}
    \caption{JavaScript deobfuscation process}
    \label{fig:Use-Case-js-deobfuscation-process}
    \vspace{-1ex}
\end{figure}

Upon selection, in the connection that contains the script, \GView automatically recognizes the HTTP headers, and it is capable of automatically stripping them to identify the script data.
The next step will be to analyze that data just like in the previous step. \GView will identify it as a JavaScript file type and choose a \gvcomp{Lexical Viewer} that allows us to see the code highlighted and apply certain transformations (e.g. deobfuscation steps) if needed. Figure~\ref{fig:Use-Case-js-deobfuscation-process} shows how the deobfuscation process works for scripts. In our particular case, we apply three deobfuscation layers: 1) comments are removed; 2) unescaping, and 3) concatenation of strings.

A very common technique to make static analysis more difficult is to add random comments between the lines of a JavaScript program, making it harder for someone to read and understand it. However, since the JavaScript plugin works like a parser, it communicates with the \gvcomp{Lexical Viewer} and provides this information to it. As such, \GView can remove all comments and make the code more readable (see Figure~\ref{fig:Use-Case-js-deobfuscation-process}).

The resulting file is still obfuscated. A closer look actually shows three obfuscation techniques that are being used:
\\
\textbf{Unescaped characters}: replace an existing character with its Unicode code.
\\
\textbf{String concatenation}: a string is not written directly, but as a result of a concatenation process between multiple strings. This technique avoids simple detection mechanisms that are looking for a particular string.
\\
\textbf{Reversing strings}: strings are kept reversed, and they are turned into their normal form during execution.
The JavaScript plugin possesses several callbacks that can be used by a smart viewer to fully deobfuscate a text - as shown in the last step in Figure~\ref{fig:Use-Case-js-deobfuscation-process}.

As a first clue, we now know that JavaScript provides two links: one for a picture and another one for an executable file. \GView keeps track of all additional files that were extracted from the PCAP and makes it easier to navigate backward to the PCAP file and search the image instead. Once located and opened such a file, \GView identifies it as a picture and uses a \gvcomp{Image Viewer} to show its content. As it turns out, that image contains a password.

Moving to the second link found in JavaScript and opening its content, we can see that it is identified as a Portable Executable. The information panel guides the investigation by showing a ZIP icon and the export name (sfxzip.exe) that indicates that this is a self-executable zip. 
This is important, as the code of this file is not really relevant: it just extracts some files from the archive and runs them.

\begin{figure}[htbp]
   \centering
    \includegraphics[width=0.9\textwidth]{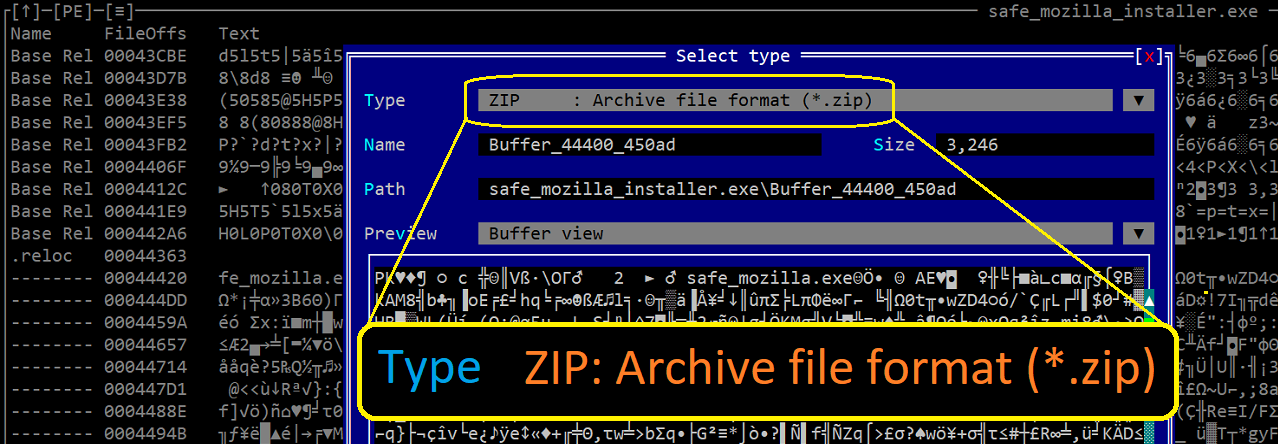}
    \vspace{-1ex}
    \caption{Automatic identification of a ZIP component}
    \label{fig:Use-Case-zip-autoidentification}
    \vspace{-1ex}
\end{figure}

The way a self-extract archive works is usually by adding the actual archive as a resource or at the end of the file. We will further use another collaboration mechanism in \GView, by selecting a part of the file (the archive part) and asking \GView to reanalyze it again. As shown in Figure~\ref{fig:Use-Case-zip-autoidentification}, \GView automatically identifies it as a zip archive and puts it into a \gvcomp{Container Viewer}, which communicates with the ZIP plugin to extract the executable ("safe\_mozila.exe").

\begin{figure}[htbp]
   \centering
    \includegraphics[width=0.9\textwidth]{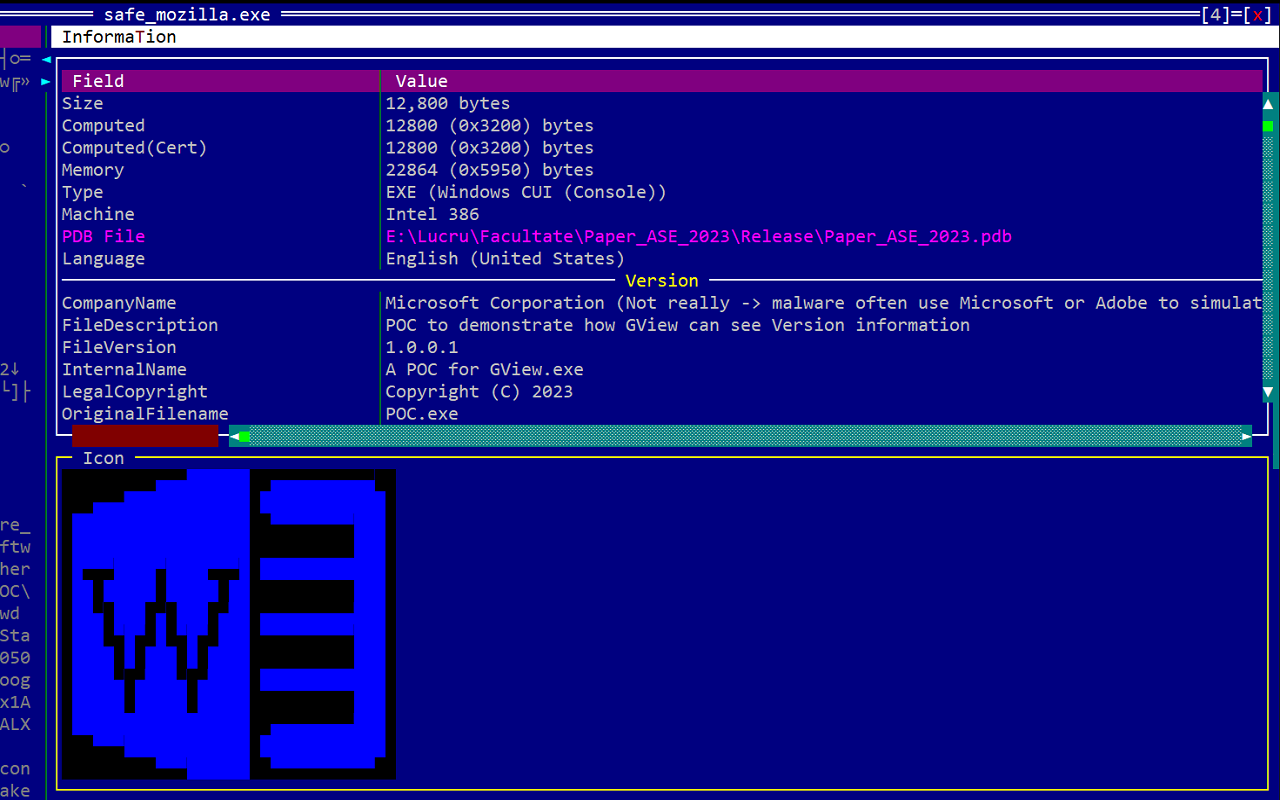}
    \caption{A fake Microsoft Word icon and version information revealed}
    \label{fig:Use-Case-fake-word-icon}
    \vspace{-1ex}
\end{figure}

Once the new binary is opened, we can immediately notice that it has an icon that looks like the one Microsoft Word uses for files, and also a version information that states that the company that has built this program is Microsoft (Figure~\ref{fig:Use-Case-fake-word-icon}). This is consistent with a technique used by attackers to trick a user into clicking something that looks like a document/folder/picture/video. 

\begin{figure}[htbp]
   \centering
    \includegraphics[width=0.9\textwidth]{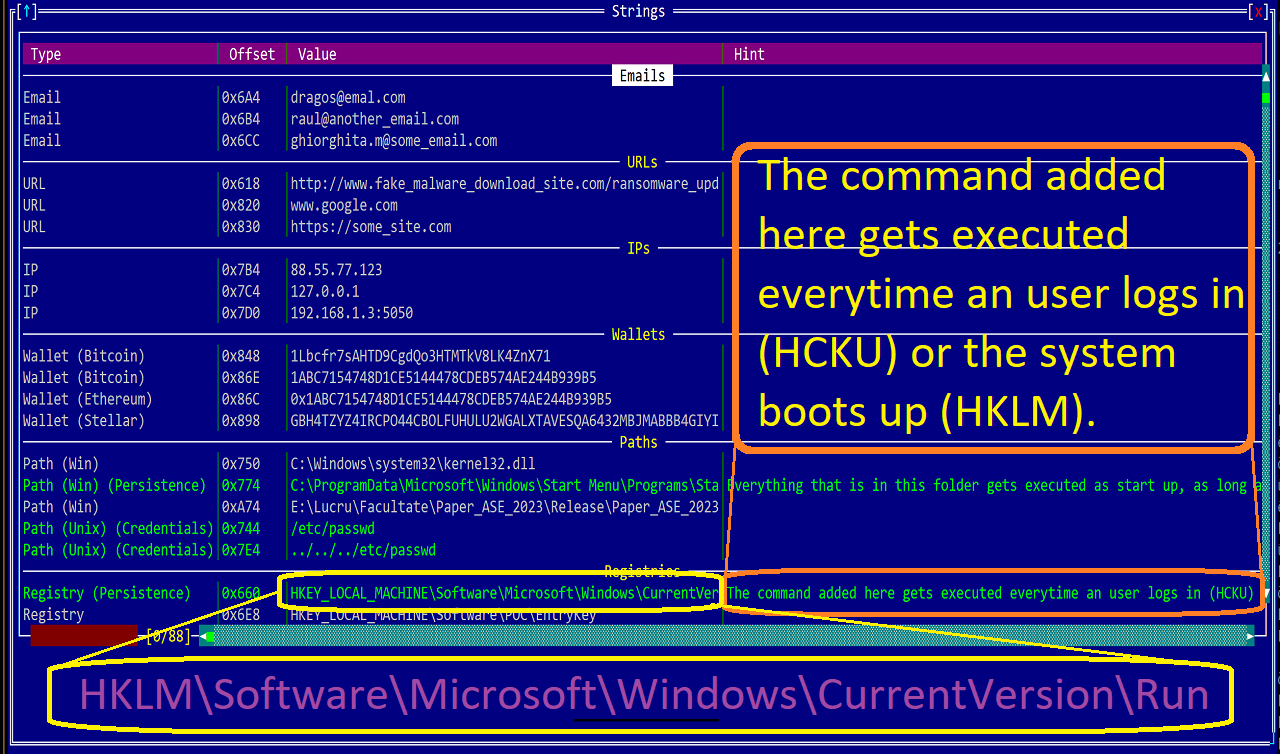}
    \vspace{-1ex}
    \caption{Artifacts automatically identified within ransomware and their meaning}
    \label{fig:Use-Case-artefacts}
    \vspace{-1ex}
\end{figure}

The next step is to use a special method from \GView that allows extracting artifacts (e.g., string/buffers that might indicate a possible malicious intent). In our case \GView identifies IP addresses, URLs, email addresses, registry keys, paths, and wallet addresses (Figure~\ref{fig:Use-Case-artefacts}). For each one of them, a risk assessment and some explications related to where that artifact can be used within an attack are provided.

\begin{figure}[htbp]
    \vspace{-2ex}
   \centering
    \includegraphics[width=0.9\textwidth]{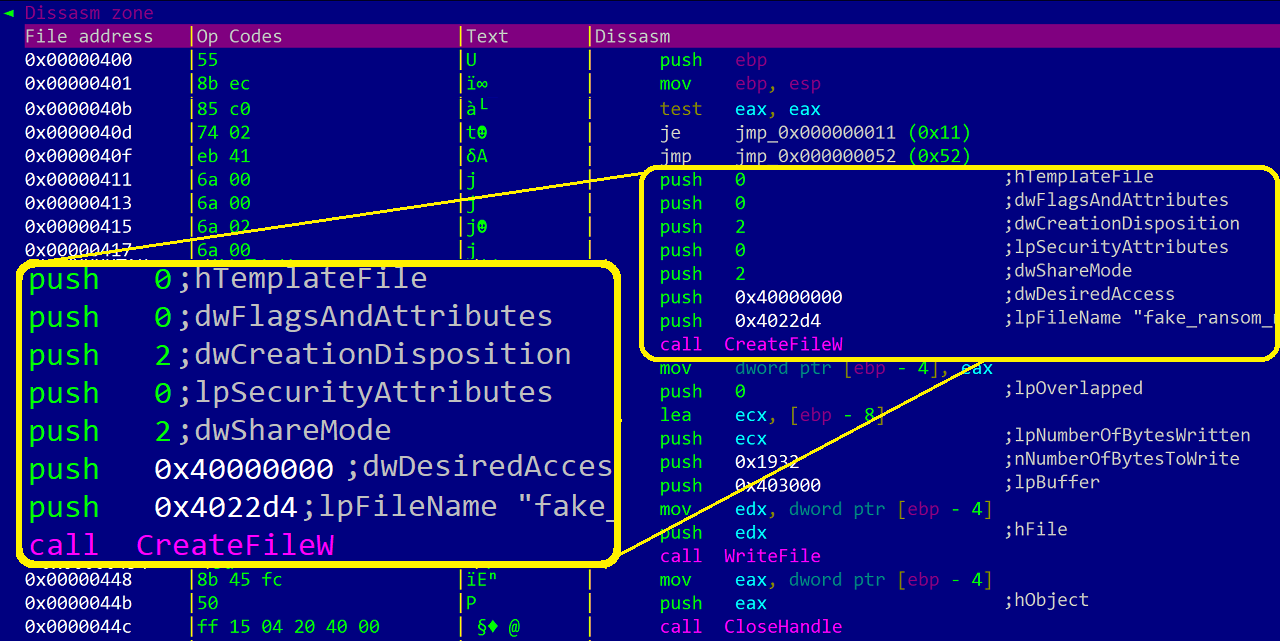}
    \vspace{-1ex}
    \caption{Code disassembly and its annotation}
    \label{fig:Use-Case-disassembly}
\end{figure}

At this point, we have extracted most of the data that reflects security hints. For more details, a deeper analysis of the code should be performed. With this in mind, a forensics engineer can use the disassembly smart viewer to better understand the behavior of that sample. As shown in Figure~\ref{fig:Use-Case-disassembly}, \GView automatically identifies several Windows APIs that are being used and also maps their parameters over the stack-based call. 

We can also easily deduce the content that is going to be written to the ransom note and, since we know its content relative to our current file offset, we can request a re-analysis of that buffer with another communication mechanism from \GView. It is often the case that ransom notes are not written directly (as text) in ransomware. To avoid detection based on this type of artifact (ransom note), a malware writer prefers to use a picture or, in our case, a form of ASCII Art that a human will understand, but an automatic system might not. \GView understands buffer encoding and can show it using another specialized smart viewer so that a security researcher can understand its meaning. 

\begin{figure}[h]
\centerline{\includegraphics[scale=0.4,interpolate]{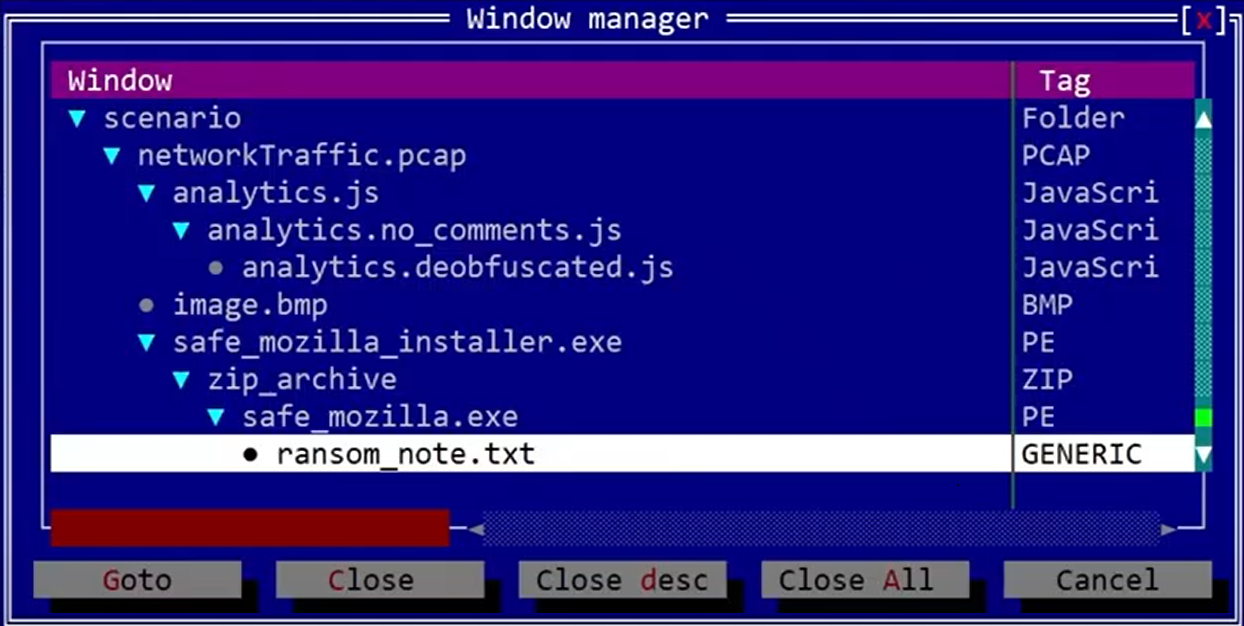}}
\caption{Scenario overview}
\label{fig:Use-Case-scenario-overview}
    \vspace{-1ex}
\end{figure}

At the end, \GView can display an overview of the entire scenario, by showing all the artifacts that the security engineer interacted with (see Figure~\ref{fig:Use-Case-scenario-overview}).

For this example, we have not used real malware, but a simulated one, to allow another researcher to replicate our analysis step without providing access to malware.

Alternatively, to analyze this scenario with multiple tools, a forensics engineer would require: a tool to parse and extract a network capture; a JavaScript analyzer \& deobfuscator; a zip extractor; an image viewer to find out the password; a resource viewer; a binary disassembly; and a text decoder/encoder.

Using \GView, all of these features are built-in, so, there is no need to use other tools or to write scripts that glue the output of one tool to the input of another. Furthermore, \GView also automatically identifies several hints that make labeling this case as malicious easier:

$\checkmark$~use of password-protected archives (with the password provided via an image) as the delivery mechanism;

$\checkmark$~usage of obfuscated JavaScript (random comments, string concatenation, unescaped characters, and reversing buffers);

$\checkmark$~final payload icon looks like the one Microsoft Office is using;

$\checkmark$~version information field states that this tool was created by Microsoft;

$\checkmark$~several artifacts related to sample persistence were found (registry key and startup path);

$\checkmark$~email addresses and wallets that indicate a possible ransom demand.

We also evaluated how \GView compares to other tools that researchers might use to accomplish the steps of this scenario. The evaluation is available in \ref{appendix:gview_tool_speed_comparison}.

\section{Evaluation and Impact}
\label{sec:impact}

We measured the impact of \GView using two methodologies: by evaluating already existing tools and also by studying its usability in a synthetic scenario that resembles real life attacks. 

We evaluated \GView against forty-one freeware tools and compared functionalities and usability (see~\ref{annex:generic_comparison}).
Our test showed that \GView improved the quality of results and reduced the analysis time by half, adding a major benefit when analyzing complex attacks that might require days/weeks to perform a full forensics investigation. 

We reexamined \GView using seven well-known tools, considering it from a binary standpoint with respect to the PE file format. Our comparison demonstrates that we package nearly all the features necessary for researchers under an MIT license, even though we do not yet have decompiler support. Appendix \ref{annex:pe_specific_comparison} goes into great detail about the comparison.

We have also evaluated \GView with the help of some security researchers from a security provider company. The evaluation consists in analyzing 100 PE (Portable Executable) samples collected over a period of two weeks. Of these samples, 50 were malicious and 50 were benign. The participants implicated in this study have different seniority levels:
\begin{itemize}
    \item junior - a person with less than 1 year of forensics experience and reverse engineering;
    \item middle - a person with 1 to 3 years of experience as a forensics engineer;
    \item senior - more than 3 years of experience as a forensics engineer.
\end{itemize}
Each one of the participants was asked to analyze the files using nothing but the \GView tool or a sandbox that runs Windows 10 and exports behavioral actions. Their task was to assign each file a verdict: malware or benign. For every sample in the data set,  two evaluations were submitted: one based on a sandbox report and one based on the hints and information provided by the \GView tool. 

We have established a set of rules, on how the sandbox report should be analyzed, aiming to reduce as much possible the bias (what you do in case of crashes, how libraries are executed, and so on.).

We have analyzed the following metrics during our experiment (See~\Cref{table:metrics_comparison}):
\begin{itemize}
    \item the time needed to analyze all samples (assuming all samples are analyzed one after another). It is important to notice that sandbox execution implies extra time needed to extract behavioral information;
    \item the accuracy in those two cases (by accuracy, we refer to the percentage of the total number of files that were correctly classified as malicious or benign);
    \item the detection rate (how many malicious files were correctly classified);
    \item the false positive rate (how many benign files were incorrectly marked as malicious).
\end{itemize}
\begin{table}[h]
\centering
\resizebox{0.95\columnwidth}{!}{
\begin{tabular}{ccccccccc}
\hline
Seniority &\multicolumn{2}{c}{Avg.time} & \multicolumn{2}{c}{Accuracy} & \multicolumn{2}{c}{Detection Rate} & \multicolumn{2}{c}{FP Rate}   \\ 
          &Sandbox&GView&Sandbox&GView&Sandbox&GView&Sandbox&GView \\
\hline
Junior & 9.5 min & 2.7 min & 82\% & 81\% & 64\% & 66\% & 0\% & 17\% \\
Middle & 8.1 min & 0.8 min & 86\% & 83\% & 72\% & 84\% & 0\% & 9\% \\
Senior & 7.8 min & 0.5 min & 87\% & 90\% & 74\% & 94\% & 0\% & 3\% \\
\hline
\end{tabular}
}
\caption{Comparison of Metrics for Sandbox and GView}
\label{table:metrics_comparison}
\end{table}

It is important to notice that the average time for running a sample in the sandbox is around 7.5 minutes. This means that after executing a sample in a sandbox, the forensic engineer only has to look at the list of events that were extracted and make a decision.

The following conclusions were drawn after this experiment:
\begin{enumerate}
    \item neither dynamic nor static analysis is sufficient for a proper analysis of a malicious sample;
    \item using a tool such as \GView can heavily improve the time needed for analysis;
    \item \GView can be used as a prefilter method for a dynamic behavior (first you try to quickly identify a file based on what \GView tool provides, and if no conclusive information is provided, use a sandbox for dynamic processing);
    \item \GView tool identifies artifacts that could be seen at a clean file and as such it is more likely to hint that a sample is malicious even if it is not.
\end{enumerate}

\section{Conclusion}
\label{sec:concl}

In this paper, we presented a versatile tool-assistant, \GView, designed to target a set of problems that a forensics engineer has when analyzing a cyber attack. Even if \GView was designed as an assistant tool, it can easily be changed to automatically perform some additional steps (e.g., hint analysis) and export their output to a file. This could be beneficial for various machine learning models that could use \GView's output to train a model. The same logic could be applied to LLM models, where \GView output could be seen as input source.

Future work will focus on adding more plugins for various data types (e.g., IPA - iPhone format, DEX - Android Java Binary, etc.), and generic integration with emulators for various architectures (e.g., MSIL, Java bytecode, Python, x86, x64, JavaScript), which will further assist with behavior hints. We also intend to incorporate support for dynamic analysis, with the mention that its use will be optional so that fast static analysis will still be possible.

\bibliographystyle{plain}
\bibliography{refs.bib}

\clearpage
\appendix

\section{Generic Feature Comparison}
\label{annex:generic_comparison}

In this section, we evaluate forty-one freeware tools (described in Table~\ref{tab:legend_tools}) against \GView features. 
Our evaluation in Table~\ref{tab:generic_tools_comarison} focuses on the following features, that are required for a tool to be useful in forensics investigations:
    \\
    \textbf{OS support} - Capability to run on Windows, Linux, Mac or through a SSH console;
    \\
    \textbf{GUI/TUI} - Whether the tool has a visual interface: a graphical or a textual one;
    \\
    \textbf{CMD line support} - Capability to run cmd line commands;
    \\
    \textbf{Automatic artifact identification} - Capability to recognize and highlight intermediary products (other PE, other types);
    \\
    \textbf{Hashing} - Capability to view some hash results based on hashing parts of data;
    \\
    \textbf{Graphical reps. entropy} - Capability to display visual representation of the entropy;
    \\
    \textbf{Results Export (CSV, XML, HTML, etc.)} - Capability to extract intermediary artifacts;
    \\
    \textbf{String extraction ASCII Unicode} - Capability to recognize strings: ASCII and Unicode;
    \\
    \textbf{String identification - URL, IP, base64, etc.} - Capability to identify specific strings and categorize them;
    \\
    \textbf{Scanner for suspicious artifacts} - Capability to recognize that some artifacts are suspicious; for instance some registry keys that should not be used;
    \\
    \textbf{Hex Viewer (Basic)} - Capability for a hex view: showing binary bytes along their hexadecimal value;
    \\
    \textbf{Function name demangler} - Capability to demangle exported functions from libraries;
    \\
    \textbf{Digital Signature} - Capability to identify digital signature;
    \\
    \textbf{Hex Viewer (Color highlighting)} - Capability to highlight relevant information depending on the zone or relevance;
    \\
    \textbf{Flows - reinterpret data parts} - Capability that some selected data or artifact can be directly reinterpreted into the same tool;
    \\
    \textbf{Binary compare} - Capability to compare two binary data to show differences;
    \\
    \textbf{Container Archive Extractor} - Capability to extract intermediary artifacts from archives;
    \\
    \textbf{Dissasm support} - Capability to disassembly code data.

As seen in Table~\ref{tab:generic_tools_comarison}, most tools are designed with a single purpose (e.g., extract a macro-VBA from a document) and as such useless for other scenarios.

\begin{table}[H]
\resizebox{1\columnwidth}{!}{
\begin{tabularx}{1.1\textwidth}{llllllll}
\toprule
Tool&T1&T2&T3&T4&T5&T6&T7\\
\midrule
Name&TrID&Detect-It-Easy&ExifTool&DroidLysis&zipdump.py&msitools (msiinfo)&strings.py\\
License&Free, unknown&MIT&GPL&MIT&Public Domain&LGPL 2.1 or later&Public Domain\\
Version&2.24&3.08&12.64&3.4.5&0.0.20&0.102&0.0.2\\
\bottomrule
\end{tabularx}
}
\\
\resizebox{1\columnwidth}{!}{
\begin{tabularx}{1.1\textwidth}{llllllll}
\toprule
Tool&T8&T9&T10&T11&T12&T13&T14\\
\midrule
Name&disitool&signsrch&UniversalExtractor2&7-Zip&wxHexEditor&bulk\_extractor&Hachoir\\
License&Public Domain&Free, unkown&GPL 2.0&LGPL&GPL 2.0&Public Domain, MIT&GPL 2.0\\
Version&0.4&0.2.4&9.53.0.20230629&23.01&0.23&2.0.3&3.2.0\\
\bottomrule
\end{tabularx}
}
\\
\resizebox{1\columnwidth}{!}{
\begin{tabularx}{1.1\textwidth}{llllllll}
\toprule
Tool&T15&T16&T17&T18&T19&T20&T21\\
\midrule
Name&file-magic.py&StringSifter&PEframe&dllcharacteristics.py&PEFile&PE Tree&pedump\\
License&Public Domain&Apache 2.0&Free, unknown&GPL 3.0&MIT&Apache 2.0&MIT\\
Version&0.0.5&3.20230711&6.1.0&undefined&2023.2.7&1.0.29&0.6.6\\
\bottomrule
\end{tabularx}
}
\\
\resizebox{1\columnwidth}{!}{
\begin{tabularx}{1.1\textwidth}{llllllll}
\toprule
Tool&T22&T23&T24&T25&T26&T27&T28\\
\midrule
Name&pecheck&PortEx&CyberChef&Cutter&capa&010 editor&Bytecode Viewer\\
License&Public Domain&Apache 2.0&Apache 2.0&GPL 3.0&Apache 2.0&Comercial product&GPL 3.0\\
Version&0.7.10&4.0.0&10.5.2&13.0.2&5.1.0.20230418&13.0.2&2.11.2\\
\bottomrule
\end{tabularx}
}
\\
\resizebox{1\columnwidth}{!}{
\begin{tabularx}{1.12\textwidth}{lllllll}
\toprule
Tool&T29&T30&T31&T32&T33&T34\\
\midrule
Name&Dependecy walker&EXEINFO PE&Hollows Hunter&PE-bear&file&PEiD\\
License&Free, unknown&Commercial&BSD-2-Clause&GPL 2.0&BSD 2-Clause Alike&Unkown\\
Version&2.11.2&0.0.7.20221209&0.0.7.20221209&0.6.5.20230308&5.45&0.95.0.20221115\\
\bottomrule
\end{tabularx}
}
\\
\resizebox{1\columnwidth}{!}{
\begin{tabularx}{1.1\textwidth}{llllllll}
\toprule
Tool&T35&T36&T37&T38&T39&T40&T41\\
\midrule
Name&pe-sieve&pestudio&Yara&bearparser&IdaFree&Hiew32Demo&iaito
\\
License&BSD-2-Clause&Standard free&BSD-3-Clause&BSD 2-Clause&Comercial&Comercial&GPL 3.0\\
Version&0.3.6&9.53.0.20230629&4.3.2&0.3&7.6.210526&Hiew8 DEMO&5.8.8\\
\bottomrule
\end{tabularx}
}
\\

\centering
\caption{Legend for the Table~\ref{tab:generic_tools_comarison}\label{tab:legend_tools}}
\end{table}

\begin{table}[H]
\centering

\scriptsize

\resizebox{1.0\columnwidth}{!}{
\begin{tabularx}{1.25\textwidth}{llllllllllllllllllllll}
\toprule
Category&T1&T2&T3&T4&T5&T6&T7&T8&T9&T10&T11&T12&T13&T14&T15&T16&T17&T18&T19&T20&T21\\
\midrule
Support for: Windows&\checkmark&\checkmark&\checkmark&\checkmark&\checkmark&-&\checkmark&\checkmark&\checkmark&\checkmark&\checkmark&\checkmark&\checkmark&\checkmark&\checkmark&\checkmark&\checkmark&\checkmark&\checkmark&\checkmark&\checkmark\\
Support for: Linux&\checkmark&\checkmark&\checkmark&\checkmark&\checkmark&\checkmark&\checkmark&\checkmark&\checkmark&-&\checkmark&\checkmark&\checkmark&\checkmark&\checkmark&\checkmark&\checkmark&\checkmark&\checkmark&\checkmark&\checkmark\\
Support for: Mac&-&\checkmark&\checkmark&\checkmark&\checkmark&\checkmark&\checkmark&\checkmark&\checkmark&-&\checkmark&\checkmark&\checkmark&\checkmark&\checkmark&\checkmark&\checkmark&\checkmark&\checkmark&\checkmark&\checkmark\\
Support for: SSH&\checkmark&\checkmark&\checkmark&\checkmark&\checkmark&\checkmark&\checkmark&\checkmark&\checkmark&?&\checkmark&-&\checkmark&\checkmark&\checkmark&\checkmark&\checkmark&\checkmark&\checkmark&\checkmark&\checkmark\\
GUI/TUI&-&\checkmark&-&-&-&-&-&-&-&\checkmark&\checkmark&\checkmark&-&\checkmark&-&-&-&-&-&\checkmark&-\\
CMD line support&\checkmark&\checkmark&\checkmark&\checkmark&\checkmark&\checkmark&\checkmark&\checkmark&\checkmark&\checkmark&\checkmark&-&\checkmark&\checkmark&\checkmark&\checkmark&\checkmark&\checkmark&\checkmark&-&\checkmark\\
Automatic artifact identification&\checkmark&\checkmark&\checkmark&\checkmark&-&-&-&-&\checkmark&\checkmark&-&-&-&\checkmark&\checkmark&-&\checkmark&-&\checkmark&\checkmark&\checkmark\\
Hashing&-&\checkmark&\checkmark&\checkmark&-&-&-&-&-&-&-&-&-&-&-&-&\checkmark&-&\checkmark&\checkmark&-\\
Graphical reps. entropy&-&\checkmark&-&-&-&-&-&-&-&-&-&-&-&-&-&-&-&-&\checkmark&-&-\\
Results Export (CSV, XML, HTML, etc)&-&\checkmark&\checkmark&\checkmark&\checkmark&-&-&-&-&\checkmark&-&-&\checkmark&-&\checkmark&-&\checkmark&-&-&-&\checkmark\\
String extraction ASCII Unicode&-&\checkmark&-&-&-&-&\checkmark&-&-&-&-&-&\checkmark&\checkmark&-&\checkmark&\checkmark&-&\checkmark&-&\checkmark\\
String identification - URL, IP, base64, etc&-&-&-&\checkmark&-&-&-&-&-&-&-&-&\checkmark&\checkmark&-&-&-&-&-&-&-\\
Scanner for suspicious artifacts&-&\checkmark&-&-&-&-&-&-&\checkmark&-&-&-&-&\checkmark&-&-&\checkmark&-&\checkmark&\checkmark&\checkmark\\
Hex Viewer (Basic)&-&\checkmark&-&-&-&-&-&-&-&-&-&\checkmark&-&\checkmark&-&-&-&-&-&\checkmark&\checkmark\\
Function name demangler&-&\checkmark&-&-&-&-&-&-&-&-&-&-&-&-&-&-&-&-&-&-&-\\
Digital Signature&-&-&-&\checkmark&-&-&-&\checkmark&-&-&-&-&-&-&-&-&\checkmark&-&\checkmark&-&\checkmark\\
Hex Viewer (Color highlighting)&-&-&-&-&-&-&-&-&-&-&-&\checkmark&-&-&-&-&-&-&-&-&-\\
Flows - reinterpret data parts&-&-&-&-&-&-&-&-&-&-&-&-&-&-&-&-&-&-&-&-&-\\
Binary compare&-&-&-&-&-&-&-&-&-&-&-&-&-&-&-&-&-&-&-&-&-\\
Container Archive Extractor&-&-&-&\checkmark&\checkmark&\checkmark&-&-&-&\checkmark&\checkmark&-&-&-&-&-&-&-&-&-&-\\
Dissasm support&-&\checkmark&-&-&-&-&-&-&-&-&-&\checkmark&-&-&-&-&-&-&-&-&-\\
\bottomrule
\end{tabularx}
}
\\
\resizebox{1.0\columnwidth}{!}{
\begin{tabularx}{1.35\textwidth}{llllllllllllllllllllll}
\toprule
Category&T22&T23&T24&T25&T26&T27&T28&T29&T30&T31&T32&T33&T34&T35&T36&T37&T38&T39&T40&T41&GView\\
\midrule
Support for: Windows&\checkmark&\checkmark&\checkmark&\checkmark&\checkmark&\checkmark&\checkmark&\checkmark&\checkmark&\checkmark&\checkmark&\checkmark&\checkmark&\checkmark&\checkmark&\checkmark&\checkmark&\checkmark&\checkmark&\checkmark&\checkmark\\
Support for: Linux&\checkmark&\checkmark&\checkmark&\checkmark&\checkmark&\checkmark&\checkmark&-&-&-&\checkmark&\checkmark&-&-&-&\checkmark&\checkmark&\checkmark&-&\checkmark&\checkmark\\
Support for: Mac&\checkmark&\checkmark&\checkmark&\checkmark&\checkmark&\checkmark&\checkmark&-&-&-&\checkmark&\checkmark&-&-&-&\checkmark&\checkmark&\checkmark&-&\checkmark&\checkmark\\
Support for: SSH&\checkmark&\checkmark&-&?&\checkmark&?&?&-&-&-&?&\checkmark&-&-&-&\checkmark&\checkmark&?&-&?&\checkmark\\
GUI/TUI&-&-&\checkmark&\checkmark&-&\checkmark&\checkmark&\checkmark&\checkmark&-&\checkmark&-&\checkmark&-&\checkmark&-&-&\checkmark&\checkmark&\checkmark&\checkmark\\
CMD line support&\checkmark&\checkmark&-&-&\checkmark&\checkmark&-&\checkmark&-&\checkmark&-&\checkmark&\checkmark&\checkmark&\checkmark&\checkmark&\checkmark&\checkmark&-&\checkmark&\checkmark\\
Automatic artifact identification&\checkmark&-&\checkmark&\checkmark&\checkmark&\checkmark&\checkmark&-&\checkmark&\checkmark&?&\checkmark&-&\checkmark&\checkmark&?&-&?&?&\checkmark&\checkmark\\
Hashing&-&-&\checkmark&\checkmark&\checkmark&\checkmark&-&\checkmark&\checkmark&-&\checkmark&-&-&-&\checkmark&\checkmark&-&\checkmark&?&\checkmark&\checkmark\\
Graphical reps. entropy&-&-&\checkmark&-&-&-&-&-&-&-&-&-&\checkmark&-&\checkmark&\checkmark&-&?&?&-&\checkmark\\
Results Export (CSV, XML, HTML, etc)&-&-&\checkmark&\checkmark&\checkmark&\checkmark&?&-&\checkmark&\checkmark&-&-&-&\checkmark&\checkmark&?&-&?&?&\checkmark&\checkmark\\
String extraction ASCII Unicode&-&-&\checkmark&\checkmark&-&-&\checkmark&-&\checkmark&?&-&-&\checkmark&?&\checkmark&-&-&\checkmark&?&\checkmark&\checkmark\\
String identification - URL, IP, base64, etc&-&-&\checkmark&-&-&-&-&-&\checkmark&?&-&-&\checkmark&?&\checkmark&-&-&-&?&-&\checkmark\\
Scanner for suspicious artifacts&-&-&\checkmark&\checkmark&\checkmark&-&\checkmark&-&\checkmark&\checkmark&\checkmark&-&\checkmark&\checkmark&\checkmark&\checkmark&-&\checkmark&?&\checkmark&\checkmark\\
Hex Viewer (Basic)&-&-&\checkmark&\checkmark&-&\checkmark&\checkmark&-&\checkmark&-&\checkmark&-&\checkmark&-&-&-&-&\checkmark&\checkmark&\checkmark&\checkmark\\
Function name demangler&-&-&-&\checkmark&\checkmark&-&-&\checkmark&-&-&?&-&?&-&-&?&-&\checkmark&?&\checkmark&\checkmark\\
Digital Signature&-&-&-&\checkmark&-&\checkmark&-&-&\checkmark&-&\checkmark&-&-&-&\checkmark&?&-&?&?&\checkmark&\checkmark\\
Hex Viewer (Color highlighting)&-&-&-&\checkmark&-&\checkmark&\checkmark&-&-&-&-&-&-&-&-&-&-&\checkmark&-&\checkmark&\checkmark\\
Flows - reinterpret data parts&-&-&\checkmark&\checkmark&-&-&-&-&-&-&-&-&-&-&-&-&-&?&?&?&\checkmark\\
Binary compare&-&-&-&-&-&\checkmark&-&-&-&-&\checkmark&-&-&-&-&?&-&?&?&-&\checkmark\\
Container Archive Extractor&-&-&\checkmark&-&-&-&-&-&\checkmark&-&-&-&-&-&-&?&-&?&?&-&\checkmark\\
Dissasm support&-&-&\checkmark&\checkmark&-&\checkmark&\checkmark&-&\checkmark&-&\checkmark&-&\checkmark&-&-&?&-&\checkmark&\checkmark&\checkmark&\checkmark\\
\bottomrule

\end{tabularx}
}

\caption{Table generic PE comparison}
\label{tab:generic_tools_comarison}

\end{table}

\clearpage

\section{File types and logs that provide evidence of an attack}
\label{annex:File_types_and_logs}

A cyber-security attack is characterized by a combination of files, payloads, and network actions rather than a single file. Because of this, assessing an attack of this nature is a difficult undertaking requiring a thorough comprehension of the different file formats, payloads, and network protocols that may be employed in these situations. The file types listed in the table below are either directly used in an attack or are extra files or security logs that a security researcher needs.

\vspace{-1ex}
\begin{table}[H]
\small
  \centering
  \begin{tabular}{|p{3cm}|p{8.5cm}|}
    \hline
    \textbf{File type} & \textbf{Description} \\
    \hline
    Binary file (.exe, .dll) &  Used by the attacker to execute code on the targeted machine \\
    \hline
    Archives (.zip, .rar) &  Used either as a transportation method for exfiltrated data \\
    \hline
    Documents (.docx, .pdf) &  Used for initial access (either as a scam or as an exploit or part of a file-less attack) \\
    \hline
    Macros (.vba) &  Used as a stage for cases where the initial access relies on Office documents that will execute a second payload \\
    \hline
    Scrips (.js, .ps1) &  Used for various type of file-less attacks \\
    \hline
    Bash (.bat, .sh) &  Used as part of execution flow or file-less attacks \\
    \hline
    Images (.png, .jpg) &  Used to convey information from the attacker that should not be easily readable pragmatically (such as a ransom note or a password to open an archive, etc.) \\
    \hline
    Links (.pif, .lnk) &  Used for lateral movement (e.g. powershell-based attacks) or partial persistence (e.g. links on desktop) \\
    \hline
    \textbf{Log type} & \textbf{Description} \\
    \hline
    Packet captures (.pcap) &  Used to analyze the network packet data \\
    \hline
    Config files (.json, .ini) &  Used to analyze configuration for various tools (they may contain indicators of miss-configurations or for malicious programs may contain C\&C addresses, passwords, wallets, etc.) \\
    \hline
    Registry hives&  Used to evaluate things like persistence, debug settings, etc. \\
    \hline
    Memory dumps &  Useful to identify vulnerabilities or credential access\\
    \hline
    System logs &  Used to identify various actions that happened on a system (login/logout information, etc.)\\
    \hline
  \end{tabular}
  \caption{File types and logs that provide evidence of an attack
  }
  \label{tab:file_types}
\vspace{-2ex}
\end{table}

\clearpage
\section{Feature Comparison For PE Files}
\label{annex:pe_specific_comparison}

In this section, we evaluate \GView with against six tools (Table~\ref{pe_specific_comparison}) that are specifically designed to extract meaningful information for PE files. The list of meaningful (security wise) information used for this evaluation includes: 
    \\
    \textbf{General information and} - Capability to read and extract information from the PE format
    \\
    \textbf{capability to recognize the corresponding PE component} - headers, sections, directories, resources, certificates, TLS sections, version info imports, relocations, overlay from the PE
    \\
    \textbf{String identification} - Capability to recognize string in the PE
    \\
    \textbf{Unicode identification} - Capability to recognize Unicode strings in PE
    \\
    \textbf{URL identification} - Capability to recognize that some strings are URLs in the PE
    \\
    \textbf{Zones Color highlighting} - Capability to highlight relevant information depending on the zone or relevance
    \\
    \textbf{Hex View} - Capability for a hex view: showing binary bytes along their hexadecimal value 
    \\
    \textbf{Flows - reinterpret data parts} - Capability that some selected data or artifact can be directly reinterpreted into the same tool
    \\
    \textbf{Automatic artifact identif.} - Capability to recognize and highlight intermediary products: other PE, other types
    \\
    \textbf{String extraction list} -  Capability to extract and collect all strings found in the PE
    \\
    \textbf{Binary compare} - Capability to compare two binary data to show differences
    \\
    \textbf{Decompiler support} -  Capability to decompilation process
    \\
    \textbf{Deobfuscation support} -  Capability to deobfuscate (different methods to increase the code redability that is hidden by authors with diverse obfuscation methods) for example: having a simple XOR function
    \\
    \textbf{Dissasm support} -  Capability to disassemble the code sections, whether it includes and recognizes Windows API functions, if it is capable of disassembling x86 or x64 languages
    \\
    \textbf{MSIL (Metadata)} -  Capability to recognize MSIL metadata
    \\
    \textbf{Icon} -  Capability to recognize and view the ICON
    \\
    \textbf{Hashing} -  Capability to run hashing functions, to view hashed values

\begin{table}[H]
    \centering 
    \resizebox{1.0\columnwidth}{!}{%
    \begin{tabularx}{1.4\textwidth}{lllllllll}
        \hline
        Tool&Detect-It-Easy&010 editor&EXEINFO PE&pestudio&PE-bear&Cutter&CyberChef&GView\\
        \hline
        License&MIT&Commercial&Commercial&Standard version free&GPL 2.0 license&GPL 3.0&Apache 2.0&MIT\\
        Tool version&3.08&13.0.2&0.0.7.20221209&9.53.0.20230629&0.6.5.20230308&13.0.2&13.0.2&0.297.0\\
        \hline
        General information&\checkmark&\checkmark&\checkmark&\checkmark&\checkmark&\checkmark&-&\checkmark\\
        Sections&\checkmark&\checkmark&\checkmark&\checkmark&\checkmark&\checkmark&-&\checkmark\\
        Directories&\checkmark&\checkmark&\checkmark&\checkmark&\checkmark&\checkmark&-&\checkmark\\
        Headers&\checkmark&\checkmark&\checkmark&\checkmark&\checkmark&\checkmark&-&\checkmark\\
        Resources&\checkmark&\checkmark&\checkmark&\checkmark&\checkmark&\checkmark&-&\checkmark\\
        String identification&\checkmark&-&\checkmark&\checkmark&-&\checkmark&\checkmark&\checkmark\\
        Unicode identification&\checkmark&-&\checkmark&\checkmark&-&\checkmark&\checkmark&\checkmark\\
        URL identification&-&-&\checkmark&\checkmark&-&\checkmark&\checkmark&\checkmark\\
        Certificates&\checkmark&-&\checkmark&\checkmark&\checkmark&?&?&\checkmark\\
        TLS sections&\checkmark&\checkmark&\checkmark&\checkmark&\checkmark&\checkmark&-&\checkmark\\
        Zones Color highlighting&-&\checkmark&-&-&-&\checkmark&-&\checkmark\\
        Hex View&\checkmark&\checkmark&\checkmark&-&\checkmark&\checkmark&\checkmark&\checkmark\\
        Container Archive extract&-&-&-&-&-&-&\checkmark&\checkmark\\
        Flows - reinterpret data parts&-&-&-&-&-&-&\checkmark&\checkmark\\
        Automatic artifact identif.&\checkmark&\checkmark&\checkmark&\checkmark&?&?&\checkmark&\checkmark\\
        String extraction list&\checkmark&-&\checkmark&\checkmark&-&\checkmark&\checkmark&\checkmark\\
        Binary compare&-&\checkmark&-&-&\checkmark&-&-&\checkmark\\
        Decompiler support&-&-&-&-&-&\checkmark&-&-\\
        Deobfuscation support&-&\checkmark&\checkmark&?&-&?&\checkmark&\checkmark\\
        Dissasm support&\checkmark&\checkmark&\checkmark&-&\checkmark&\checkmark&\checkmark&\checkmark\\
        Dissasm - Windows API&-&-&-&-&\checkmark&\checkmark&-&\checkmark\\
        Dissasm - x86 support&\checkmark&\checkmark&\checkmark&-&\checkmark&\checkmark&\checkmark&\checkmark\\
        Dissasm - x64 support&\checkmark&\checkmark&\checkmark&-&\checkmark&\checkmark&\checkmark&\checkmark\\
        MSIL (Metadata)&\checkmark&?&\checkmark&\checkmark&?&?&?&\checkmark\\
        Version info&\checkmark&-&\checkmark&\checkmark&-&\checkmark&\checkmark&\checkmark\\
        Icon&?&-&\checkmark&\checkmark&\checkmark&-&-&\checkmark\\
        Hashing&\checkmark&\checkmark&-&\checkmark&\checkmark&\checkmark&\checkmark&\checkmark\\
        Entropy&\checkmark&-&-&\checkmark&-&\checkmark&\checkmark&\checkmark\\
        Imports&\checkmark&\checkmark&\checkmark&\checkmark&\checkmark&\checkmark&-&\checkmark\\
        Relocations&\checkmark&\checkmark&-&\checkmark&\checkmark&\checkmark&-&\checkmark\\
        Overlay&\checkmark&?&\checkmark&\checkmark&\checkmark&\checkmark&-&\checkmark\\
        \hline
    \end{tabularx}
}

    \caption{PE specific comparison}
    \label{pe_specific_comparison}
\end{table}

\clearpage
\onecolumn
\section{Current GView data identifiers}
\label{appendix:gview_data_identifiers}

\noindent
Table~\ref{tab:gview_data_identifiers} contains a list of all currently supported data identifiers. A complete and updated list can be found on \GView's GitHub project page. For each data identifier, we provide a:
    \\
    \textbf{Tag} - a capitalize word that identifies that data identifier (e.g. PE for Windows Binaries that use the \textit{P}ortable \textit{E}xecutable format
    \\
    \textbf{Format} - data format (text, binary, image, ...)
    \\
    \textbf{Identification} - how we identify thata of this type (could be one of: \textbf{magic} (via a special numerical value from the header), \textbf{ext} (via the extension of that file) or \textbf{heuristic} (a specific logic - e.g. regular expression)
    \\
    \textbf{Smart viewers} - a list of associated smart viewers for the current data identifier
    \\
    \textbf{Description} - a description of that data identifier

\begin{table}[H]
  \resizebox{1.0\columnwidth}{!}{
  \begin{tabularx}{1.1\textwidth}{l|l|l|l|X}
    \hline
    \textbf{Tag} & \textbf{Format} & \textbf{Identification} & \textbf{Smart viewers} &\textbf {Description} \\
    \hline
    BMP & image & magic & ImageView, BufferView & Support to show bitmap images (*.bmp) \\
    CPP & text & ext, heuristic & LexicalView, TextView, BufferView & C++ headers and source files \\
    CSV & text & ext & TableView, TextView, BufferView & for Comma Separated Values (CSV) or Tab Separated values (TSV) data sheet \\
    ELF & binary & magic & BufferView & Linux binary files \\
    ICO & image & magic & ImageView, BufferView & for icon and cursor files (including PNG formats) \\
    INI & text & ext, heuristic & LexicalView, TextView, BufferView & for initialization/configuration files or TOML (Tom's Obvious Minimal Language) markup files \\
    ISO & binary & magic & ContainerView, BufferView & for ISO Archives (based on ECMA 119\footnote{\url{https://www.ecma-international.org/wp-content/uploads/ECMA-119_4th_edition_june_2019.pdf}}) \\
    JOB & binary & magic & BufferView & for job/schedule windows task files \\
    JS & text & ext, heuristic & LexicalView, TextView, BufferView & JavaScript/TypeScript analyzer and deobfuscator \\
    JSON & text & ext, heuristic & LexicalView, TextView, BufferView & JSON file format parser \\
    JT & text & text & TextView, BufferView & for Jupiter Tessellation files (3D data format - ISO 14306:2012\footnote{\url{https://www.iso.org/standard/60572.html}}) \\
    LNK & binary & magic & BufferView & for LNK (Link) files used in Windows to simulate a symbolic link \cite{thabet2011stuxnet} \\
    MACHO & binary & magic & BufferView & MAC/OSX binary file format \\
    PREFETCH & binary & magic & BufferView & for prefetch files that Windows uses to cache applications that were executed (often used in forensics to identify tools that were used during the last couple of weeks). \\
    MAM & binary & magic & BufferView & for compressed Prefetch files (compression method is Microsoft XPRESS Huffman \footnote{\url{https://learn.microsoft.com/en-us/openspecs/windows\_protocols/ms-xca/a8b7cb0a-92a6-4187-a23b-5e14273b96f8}} (LZXPRESS)) \\
    PCAP & binary & magic & ContainerView, BufferView & for files that contain a network packet capture \\
    PE & binary & magic & BufferView, DisasmView & Windows Portable Executable (PE) binaries (*.exe, *.dll, *.sys) \\
    PYEXTRACTOR & binary & magic & BufferView & for PyInstaller files (CArchive format \footnote{\url{https://pyinstaller.org/en/latest/advanced-topics.html\#carchive}}) \\
    VBA & text & ext, heuristics & LexicalView, TextView, BufferView & Macros: VBA (Visual Basic for Application) and VBS (Visual Basic Scripts) \\
    ZIP & binary & magic & ContainerView, BufferView & ZIP unpacker \\
    \hline
    FOLDER & folder & - & ContainerView & Generic visualization for the content of a folder \\
    TEXT & text & - & TextView, BufferView & Generic visualization for a text file/buffer \\
    BINARY & binary & - & BufferView & Generic visualization for a binary file/buffer \\
    \hline
  \end{tabularx}
  }
  \caption{List of currently available data identifies}
  \label{tab:gview_data_identifiers}
\end{table}

\clearpage
\onecolumn
\section{Tool time comparison}
\label{appendix:gview_tool_speed_comparison}

The table below showcases required time for each step in the scenario in Section~\ref{sec:UseCase}. \GView is compared with freeware software that is suited for each task. In both cases, each tool was tested from the perspective of an expert and some steps were skipped: analyzing other intermediary JavaScript files. 
In this scenario, \GView performed the task in approximately three minutes, two times faster than other tools. While it might not seem much, in practice, where an analysis takes days, weeks and even months, that efficiency comes really in hand. \GView really shined by guiding the analyst and recognizing each file type and suggesting the appropriate method to analyze it. It was also able to extract the payloads from HTTP protocol directly and open them in the most suited Smart View. 
The entire experience can be seen:
\begin{itemize}
\item how other tools behave on our scenario: \url{https://youtu.be/fRN-rXD9ZXg}
\item how \GView behave on our scenario: \url{https://youtu.be/mX3lkB_tf3w}
\item screencast with a detailed scenario for \GView: \url{https://youtu.be/Z1nTRKPewCg}
\end{itemize}

\begin{table}[H]
\begin{tabularx}{0.78\textheight}{lllll}
\toprule
Step&Substep&Tool&ToolTime (sec)&GViewTime (sec)\\
\midrule
PCAP file analysis&Analyze traffic&Wireshark&7&5\\
Analyze suspicious scripts&Search for suspicious scripts&Wireshark&9&6\\
&Extract connection data&Wireshark&11&N/A\\
&Remove HTTP header&Notepad++&6&N/A\\
JavaScript deobfuscation&Remove comments and unescape&de4js&10&12\\
&Save intermediary state&GView&N/A&4\\
&Reverse + concatenate&Notepad++&46&14\\
&Save intermediary state&GView&N/A&5\\
Analyze image.bmp&Search for image.bmp&Wireshark&8&3\\
&Extract connection data&Wireshark&12&N/A\\
&Remove HTTP header&Notepad++&12&N/A\\
&Recognize bmp file type&Manual&2&N/A\\
&View image.bmp&Windows Photos&5&3\\
Analyze safe\_..\_intaller.exe&Search safe\_moz...exe&Wireshark&9&4\\
&Extract connection data&Wireshark&11&N/A\\
&Remove HTTP header&Notepad++&3&N/A\\
&Recognize PE file type&Manual&7&N/A\\
&View PE details&pestudio&13&9\\
&Extract overlay&pestudio&7&5\\
&Unzip overlay&7 ZIP&6&3\\
Analyze safe\_mozilla.exe&Recognize PE file type&Manual&4&N/A\\
&View PE details&pestudio&32&21\\
&View icon&Pe-Bear&15&2\\
&Analyze dissasembled code&IDA Free&28&21\\
&Extract suspicious context&IDA Free&41&20\\
&Adjust suspicious content&Notepad++&22&0\\
&View suspicious content&Notepad++&8&4\\
\bottomrule
\end{tabularx}
\\

  \caption{Time comparison for each task in the scenario mentioned above}
  \label{tab:gview_tool_comparison}
\end{table}

The test was done on a laptop machine with the following details: Windows 10 Pro, Processor 8th Generation Intel® Core™ i7-8750H, 16 GB  RAM 2666MHz DDR4, Nvidia GeForce GTX 1060 6 GB  GDDR5.

\end{document}